\documentclass[lettersize,journal]{IEEEtran}
\usepackage{amsmath,amsfonts}
\usepackage{algorithmic}
\usepackage{algorithm}
\usepackage{array}
\usepackage[caption=false,font=normalsize,labelfont=sf,textfont=sf]{subfig}
\usepackage{textcomp}
\usepackage{stfloats}
\usepackage{url}
\usepackage{verbatim}
\usepackage{graphicx}
\usepackage{cite}

\begin{document}

\onecolumn
\begin{center}
\vspace*{\fill}
{\huge This work has been submitted to the IEEE for possible publication. Copyright may be transferred without notice, after which this version may no longer be accessible.}
\vspace*{\fill}
\end{center}
\twocolumn

\newpage

\title{Helping Blind People Grasp: Enhancing a Tactile Bracelet with an Automated Hand Navigation System}

\author{Marcin Furtak*, Florian Pätzold*, Tim Kietzmann, Silke M. Kärcher, and Peter König
\thanks{* These authors contributed equally}%
\thanks{M. Furtak, F. Pätzold, T. Kietzmann, and P. König are with Institute of Cognitive Science, University of Osnabrück, 49069 Osnabrück, Germany}%
\thanks{M. Furtak and S. M. Kärcher are with feelSpace GmbH, 49074 Osnabrück, Germany}%
\thanks{P. König is with Department of Neurophysiology, University Medical Centre Hamburg-Eppendorf, 20251 Hamburg, Germany}%
\thanks{Author contributions: conceptualization, S.M.K. and P.K.; methodology, M.F., F.P., S.M.K. and P.K.; software, M.F. and F.P.; validation, M.F., and F.P.; formal analysis, M.F, and F.P.; investigation, M.F., and F.P.; resources, S.M.K. and P.K.; data curation, M.F. and F.P.; writing—original draft preparation, M.F., and F.P..; writing—review and editing, M.F., F.P., T.K., S.M.K. and P.K.; visualization, F.P.; supervision, S.M.K. and P.K.; project administration, M.F., and F.P. All authors have read and agreed to the published version of the manuscript.}%
\thanks{The data presented in this study are openly available at https://osf.io/z4kwj/ (published on 30 March 2025). The code used in this study is openly available at https://github.com/Flosener/tactile-guidance}}

\maketitle

\begin{abstract}
Grasping constitutes a critical challenge for visually impaired people. To address this problem, we developed a tactile bracelet that assists in grasping by guiding the user’s hand to a target object using vibration commands. Here we demonstrate the fully automated system around the bracelet, which can confidently detect and track target and distractor objects and reliably guide the user’s hand. We validate our approach in three tasks that resemble complex, everyday use cases. In a grasping task, the participants grasp varying target objects on a table, guided via the automated hand navigation system. In the multiple objects task, participants grasp objects from the same class, demonstrating our system’s ability to track one specific object without targeting surrounding distractor objects. Finally, the participants grasp one specific target object by avoiding an obstacle along the way in the depth navigation task, showcasing the potential to utilize our system’s depth estimations to navigate even complex scenarios. Additionally, we demonstrate that the system can aid users in the real world by testing it in a less structured environment with a blind participant. Overall, our results demonstrate that the system, by translating the AI-processed visual inputs into a reduced data rate of actionable signals, enables autonomous behavior in everyday environments, thus potentially increasing the quality of life of visually impaired people.
\end{abstract}


\section{Introduction}

\IEEEPARstart{B}{lind} people deal with a plethora of challenges that affect their quality of life on an everyday basis. Due to a lack of vision, performing some actions that normally sighted people conduct with ease proves to be very difficult, if possible at all. One such action is grasping, which is essential for more complex interactions with the environment \cite{Feix2016}, \cite{Nakamura2017}. Vision is a crucial part of the grasping process \cite{Stone2015}, both in terms of understanding the environment and planning the movement \cite{Jeannerod1999}, \cite{Maiello2021}, \cite{Smeets2019}, as well as providing feedback about the status of the process itself \cite{Schettino2003}. Therefore, visually impaired people struggle to successfully perform it even when their motor control is intact \cite{Pardhan2011}. As the size of the visually impaired population worldwide is large \cite{WHO2023} and expected to grow in the upcoming years \cite{Fricke2018}, coming up with solutions to assist them is of great interest.

Fortunately, the number of assistive devices available to the population is growing, with the ongoing development of technologies that aim to help people with visual impairment \cite{Zallio2022}. One group of solutions utilizes the principle of sensory substitution \cite{Bach-y-Rita2003}, \cite{Eagleman2023} with the idea of functionally replicating general vision using other, fully intact modalities. While successful in enabling their users to, in principle, perceive surroundings (e.g., \cite{Caraiman2019}, \cite{Gomez2014}, \cite{Hanneton2010}), these devices require prolonged training and are not easily scalable for use in more specific tasks. Another group of tools is based on sensory augmentation \cite{Macpherson2018} and intends to enhance existing or, possibly, create new senses. In that way, new functional relations between stimuli and motor actions called sensorimotor contingencies are created \cite{ORegan2001}, adding to existing sensory functions instead of replacing them. Overall, technical advancements and a better understanding of the long-term usage effects of previously developed solutions (e.g., \cite{Kaspar2014}, \cite{Brandebusemeyer2021}) enable the emergence of novel devices that will help fill the existing need for additional aid for visually impaired people \cite{WHO2022}, \cite{WHO2024}.

While some devices that aim to help with more specific tasks are already available, almost none focus on grasping. Three exemplary devices assisting in object localization and reaching – PalmSight \cite{Yu2016}, FingerSight \cite{Satpute2020}, and a tactile glove \cite{DePaz2023} – utilize tactile signals to enable scanning of the user’s surroundings. However, neither of them provides functionality to navigate the user to the specific target object, which limits their usability in a real-life scenario. Thus, developing novel tools that could aid blind people in grasping would prove highly beneficial and provide a higher chance for their usage outside the lab.

We addressed the lack of assistive grasping devices by developing a tactile bracelet specifically tailored to aid blind people in grasping \cite{Powell2024}. Our previous study experimentally validated the feasibility of utilizing tactile guiding signals compared to auditory signals. Within a simple setup of several objects placed on a shelf, participants were remotely guided to the target by an experimenter who had access to a live camera feed from the participants’ perspective. Participants could perform the task effectively using the tactile bracelet with comparable speed to auditory commands, while keeping the auditory channel unobstructed. Importantly, we received positive feedback from two blind participants regarding the bracelet's ease of use and usefulness. Overall, our results were promising and showed that upon further development, the bracelet could serve as a useful assistive device.

While the potential of the bracelet was validated successfully, several key features would need to be added to enhance its usage outside of the lab. First and foremost, the previous setup required the experimenter to send direction commands to the bracelet to successfully guide the participant’s hand instead of automated hand navigation. Secondly, as the previous study aimed to compare auditory and tactile commands to establish whether tactile guidance of the hand would be a valid solution, its setup was simple. Navigation was conducted in four general directions across two dimensions while unique target objects were placed conveniently with sparse distances between them. Subsequently, it did not tackle more challenging problems common in the more crowded and chaotic natural scenes, such as dealing with multiple instances of the same object category or handling the problem of an obstacle in the line of planned hand movement. Consequently, creating and validating a control system around the bracelet that enables its robust and autonomous usage in several demanding use cases would be an essential next step to make it usable in real-world scenarios.

In this paper, we introduce the AI-based automated hand navigation system (HANS) that enables the usage of the bracelet independently of an external operator. The core elements of the system are two object detectors working in parallel on each input frame obtained from the camera feed, one focused on detecting the objects and the other on detecting hands. The detections are then further processed by the object tracker, which keeps track of specific object instances. Additionally, a depth estimator approximates the distance from the camera to all elements in the camera's field of view. Finally, the outputs of all models serve as input for the guiding logic script that sends direction commands to the bracelet. Therefore, our system overcomes the limitation of dependence on a third person by removing the operator from the navigation control loop, thus allowing visually impaired users of the tactile bracelet to grasp selected target objects in the scene independently.

\section{Methods}

To remove the operating third person from the control loop of the navigation, we enhance the tactile bracelet with the HANS. More specifically, logic is added to the existing tactile bracelet and camera setup to integrate the visual information into the system while using AI to translate the visual stimuli into a grasping path for the user of the bracelet and consequently send vibration signals in the corresponding direction (see Figure 1 for an overview).

\subsection{Hand Navigation System}

We used the same tactile bracelet (Figure 1A) as in the study of \cite{Powell2024}. The bracelet is worn on the wrist of the participants’ dominant hand and contains four vibration motors. In this study, we tested two different types of navigation, one more complex and one simpler type. With blinfolded participants, outside of providing input at 0°, 90°, 180°, and 270° (up, right, down, and left directions, respectively) across the participants’ wrists, we interpolated guiding signals for other angles by running up to two motors at the same time with proportionally scaled intensities. In that way, we could provide guiding information for all angles from 0° to 360°, covering the whole two-dimensional space. On the other hand, in a validation of the HANS with a blind participant, we utilized the simpler navigation type: the hand would first be navigated to left or right to horizontally align the hand with the target object, before navigating up or down to also align the hand vertically with the target. Successful navigation required a fixed hand orientation with the back of the hand always on top. To collect the live video feed serving as input for the HANS, we used a small camera with a field of view of 88°, attached to experimental glasses (Figure 1B). The camera required a USB connection to the experimenter's laptop. Currently, the experimenter manually enters the target object into the system for each trial, or a list of objects is iterated automatically. However, the target selection would then be substituted by a natural language processing module in a later stage of the project.

\begin{figure}[t]
\centering
\includegraphics[width=\linewidth]{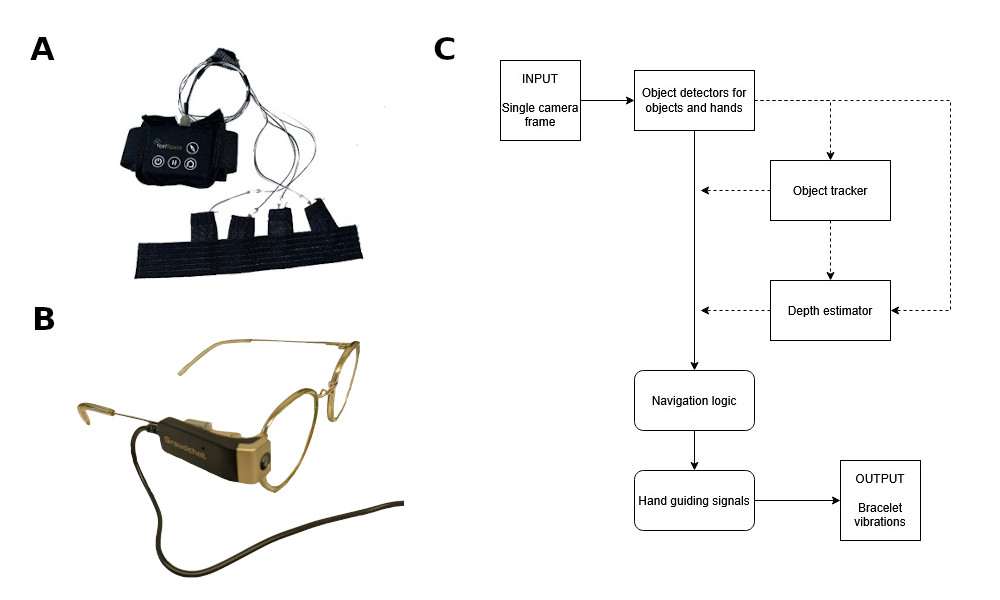}
\caption{A) The tactile bracelet. B) The camera is a lightweight device attached to experimental glasses. C) Diagram presenting the pipeline of the automated hand navigation system.}
\label{fig_1}
\end{figure}

The AI-based HANS comprises multiple components (Figure 1C). First, a live feed from the camera is passed to two vision systems that work in parallel to detect objects and hands in the scene. Relevant detections are then sent to an object tracker that performs position inference over time to deal with object occlusions. Furthermore, a depth estimator deep neural network is capable of inferring depth from a two-dimensional image to enable guidance around obstacles that are in the grasping trajectory between the hand and the target. All of these components work independently, and the object detector alone is sufficient to run the HANS, with the possibility of turning object tracking and depth estimation on or off. While this enables a more lightweight solution, all components together promise the most reliable results. Finally, the information from the components about relevant objects and the user’s hand is used to calculate a grasping trajectory that is translated into vibration commands sent to the tactile bracelet to guide the user’s hand in the corresponding direction of the target object.

\subsection{Experimental Validation}

To evaluate the ability of our system to guide hand movements autonomously, we conducted a validation study with both blindfolded and blind participants. During the experiment, participants were asked to perform a grasping task, a multiple objects task, and a depth navigation task, varying in complexity. Importantly, all scenarios were set up in contexts resembling potential real-life use cases. Hence, if validated positively, the system could be deemed robust and have the potential for further development. In the following section, we will explain the tasks and the experimental procedure in more detail.

\subsubsection{Participants}

We recruited four normally sighted participants to perform our system’s evaluation procedure. All participants provided informed consent. Each participant had prior experience with using the tactile bracelet. However, two \textit{expert} participants had extensive training for more than five hours in using the tactile bracelet with the HANS, while the other two \textit{naïve} participants had never used the HANS before but only the tactile bracelet with experimenter guidance. Additionally, we invited a blind participant for a system evaluation session. All participants were right-handed.

\subsubsection{Procedure}

\paragraph{Training}

Every participant started with the calibration of the bracelet intensity. The calibration was a simple iteration through all four motors, where the intensity would start with a baseline of 50\% for each motor. The experimenter adjusted accordingly by increasing or decreasing the vibration intensity in 5\% steps for each motor independently to compensate for individual differences in wrist anatomy and perception of tactile signals. Thus, the calibration process was used to limit the risk of skin numbness during the experiment.

Upon the successful conclusion of the calibration process, the participants were presented with training to familiarize themselves with the tactile bracelet and the navigation procedure. The training task consisted of an unrestricted number of grasping trials in which the participants could test how to interpret and use the navigation signals. If the participants self-reported the ability to use the bracelet confidently, the testing phase of the experiment started. Additionally, before the depth navigation task, the same self-paced training procedure was repeated for one target object in an unlimited number of trials, as this task introduced a new type of guiding signal.

\paragraph{Testing}

To evaluate the participants' ability to reach for the selected target object independently, we conducted a testing session that consisted of three tasks described in more detail below. In all tasks, the participants wore a blindfold and glasses with an attached camera (Figure 1B) on top of it. Further, they were seated in front of the table with several objects placed on it (Figure 2). During the experiment, all targets were selected automatically by iterating a list of objects; the experimenter manually started each task, started and concluded the trials, and rearranged the presented objects between trials when needed. Notably, the hand navigation process was fully automated and independent of the experimenter's input. Finally, the participants were instructed to perform the tasks at their own pace. We therefore used the fraction of successful trials as the primary metric of the participants' performance. Additionally, we measured the target object detection rate to assess our system’s reliability and trial durations to explore potential associations between performance and the time needed to perform the task.

\begin{figure}
\centering
\includegraphics[width=2in]{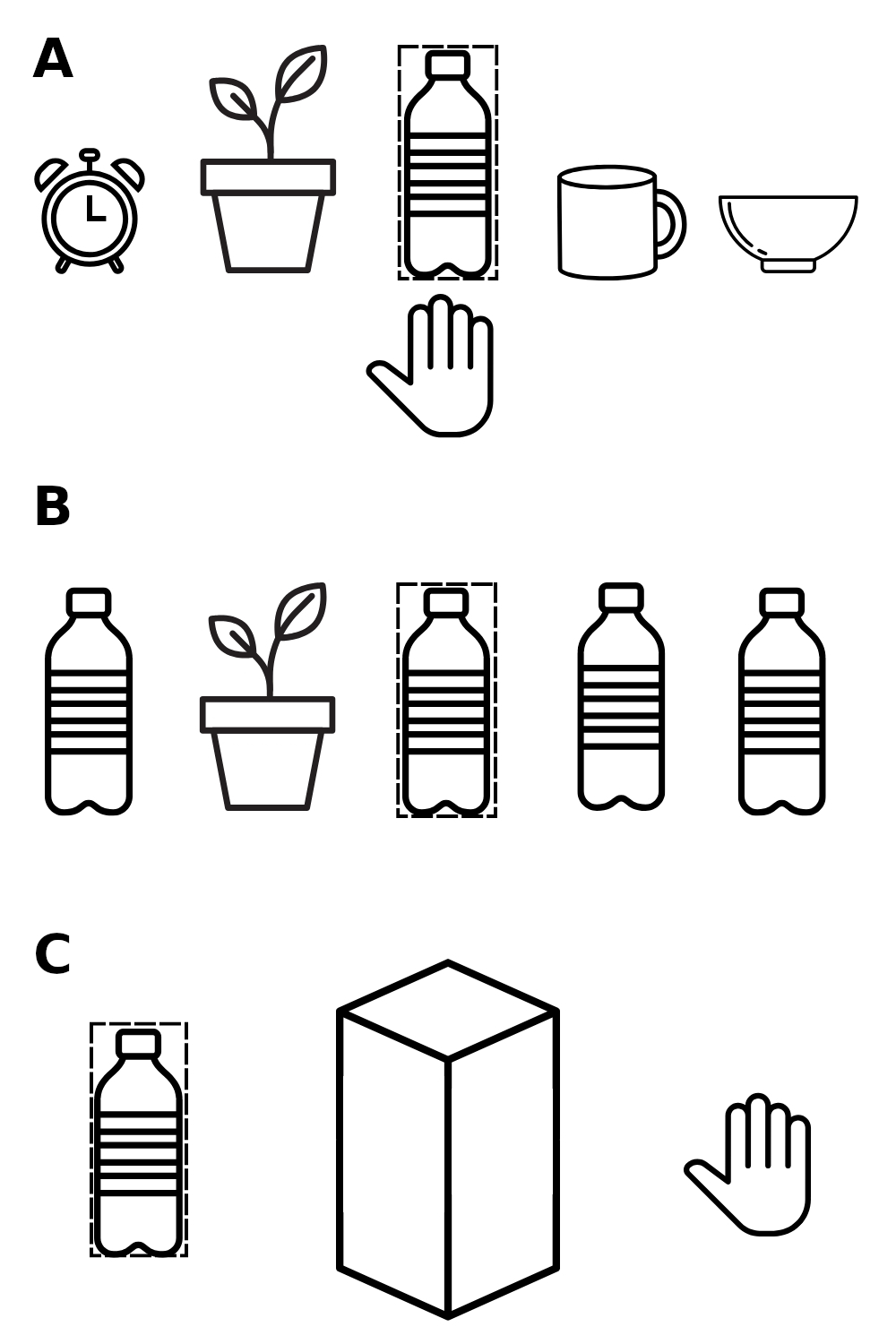}
\caption{Schematic representation of the grasping task (A), multiple objects task (B), and depth navigation task (C).}
\label{fig_2}
\end{figure}

\textbf{Grasping Task}

In the grasping task, five objects were placed next to each other on a table in a single line perpendicular to the table edge. The center object in front of the participants was placed 40 cm from the table's edge, and every object was separated by 15 cm. Each object was of a unique category; only one instance per class was presented. After the detection of both the target object and the hand, the system provided continuous guiding signals in a two-dimensional plane aimed at aligning the hand with the target. Once the hand started occluding the target object, the bounding box of the target was ‘frozen’, meaning that its last position and size were recorded and used in subsequent frames, to enable navigation once the system was unable to detect the object. After the hand reached the position in front of the target object, all of the motors vibrated simultaneously sending a pulse command, indicating that the object can be grasped by moving the hand forward (Figure 2A). Participants moved their hand back to the starting position after the trial's conclusion, and the next trial followed shortly after. In total, there were 10 trials, with the positions of the objects shuffled after five trials.

\textbf{Multiple Objects Task}

In the multiple objects task, instead of five unique objects, four instances of the same category (‘bottle’) and one of the different object categories (‘potted plant’) were placed in the scene (Figure 2B). The goal was to evaluate the system's ability to fixate on one target by utilizing the output of the object tracker. The participants were asked to grasp subsequent target objects and remove them from the scene upon grasping. If they failed to do that, the experimenter removed the intended target object. Thus, each following trial had fewer objects present in the scene. Furthermore, upon removal of all of the target objects after four trials, the position of the potted plant was changed, and the bottles were placed again in the scene by the experimenter. Again, there were 10 trials in total, with the position of the potted plant changed twice. Therefore, there were three rounds of trials with participants grasping four, again four, and then two target objects in each respective round.

\textbf{Depth Navigation Task}

Finally, in the depth navigation task, an obstacle was placed on the table, obstructing the standard hand navigation trajectory. Here, the participants started with their hand on the table on the right side of the obstacle while the target object was either on the left side or partially occluded by the obstacle. To move around the obstacle, information from the depth estimator was incorporated into the navigation logic. The obstacle map was used for determining the optimal trajectory either by sending the signal to move the hand back (subsequent double vibration of all four motors in quick succession) if the obstacle prevented horizontal movement (Figure 2C), or by navigating above the obstacle. In the task, five trials required moving the hand backward, and five trials required moving above the obstacle with the order of obstacles set up (Figure S1 in Supplemental Files) randomized for each participant.

\paragraph{Questionnaire}

After the testing session concluded, the participants were asked to complete a questionnaire related to their experience with the tactile bracelet and the HANS during the experiment. The closed section of the questionnaire consisted of 18 questions from nine topics related to the perceived comfort and ease of bracelet usage in the experimental tasks. More specifically, for each topic, two questions were formulated to assess the consistency of participants' responses. Questions were rated on a Likert scale from one (‘strongly disagree’) to five (‘strongly agree’). Additionally, three open questions related to the general experience of using the tactile bracelet and performing the training and testing sessions were included, providing participants with an option to share unrestricted feedback. For the questionnaire analysis, we aggregated the responses across participants for both questions within each topic. If one of the questions was formulated positively and the other one negatively, we would flip the scores for the negative formulation. Afterward, we calculated a median and mean response as well as the standard deviation for each topic. All the topics, questions, and corresponding scores are presented in the Supplemental Files (Table S1). Additionally, we evaluated the responses from the open questions in line with the summary statistics to gather further insights and learn more about the participants' subjective assessment of the system.

\section{Results}

For every experimental task, we divide the results into two sections. The first section covers technical aspects of one of the system components validated in that specific task. The second section presents the experimental validation results of the system obtained with the help of participants.

\subsection{Grasping a target object}
\subsubsection{Object Detection}

The underlying component for the automated hand navigation system to work is object detection. We utilize two YOLOv5 object detectors \cite{Redmon2016}, one for detecting objects and one for detecting hands in the scene (Figure 3A). This component links raw video live feed from the camera and optional components in our system that enhance reliability. Single images are pre-processed by resizing, normalization, color space conversion, and reshaping. The detections are then post-processed using non-maximum suppression. Noticeably, object detection alone is sufficient to calculate a grasping trajectory that can guide the user’s hand using the vibrations from the tactile bracelet.

To tailor the object detectors to our specific needs, both detectors were originally trained on the COCO dataset \cite{Lin2014} and later additionally retrained depending on their functionality. Firstly, we utilized the pre-trained YOLOv5 object detector network and fine-tuned it on a subset of 20 COCO categories, with 5000 instances per class each. However, as the project evolved, it benefited from a larger pool of object classes. Thus, we returned to the pre-trained model with satisfying performance to avoid further fine-tuning. More importantly, the pre-trained hand detection network was retrained on the EgoHands dataset \cite{Bambach2015}, containing 4800 images in total with four hand classes representing the perception of the hand from first-person (my\_left, my\_right) and third-person (other\_left, other\_right) perspectives. This distinction is essential as it enables successful navigation explicitly based on the user's hand location while disregarding hands from other persons in the scene. The dataset was split into training, testing, and validation with 3840, 480, and 480 images, respectively. Afterwards, we augmented the training split of the dataset by creating three versions of each image, randomly choosing from a pool of augmentations including image rotation, crop, grayscale conversion, cutout, blur, and tweaked hue, saturation, brightness, and exposure. This augmentation step resulted in 11520 training images, with a further 480 images in the test split and 480 images in the validation split of the augmented EgoHands dataset. Overall, there were 2559 instances in the my\_left class, 3418 instances in the my\_right class, 4560 instances in the other\_left class, and 4507 instances in the other\_right class.

\begin{figure}[t]
\centering
\setlength\fboxsep{0pt}
\setlength\fboxrule{0.25pt}
\fbox{\includegraphics[width=\linewidth]{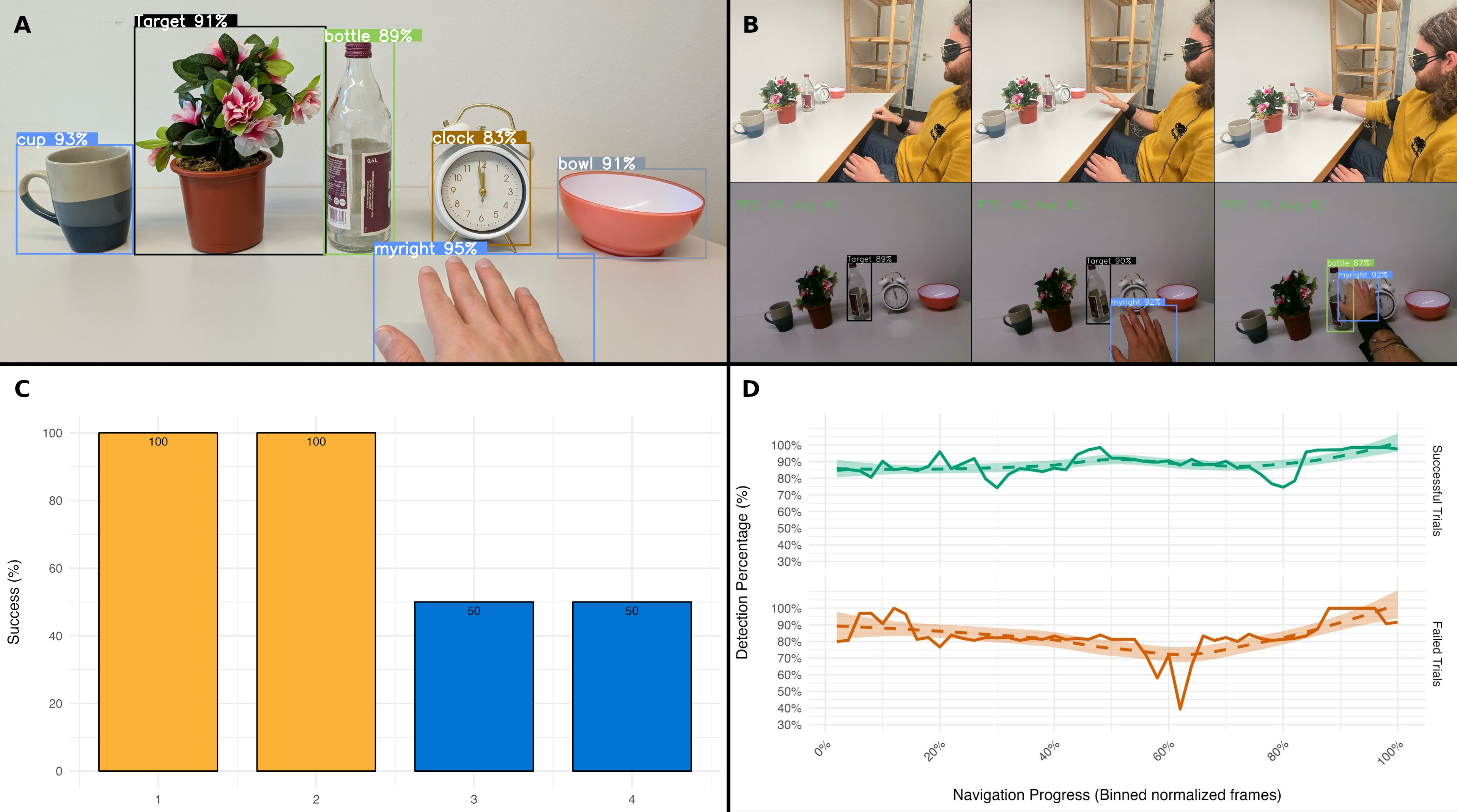}}
\caption{A) Detection of hand, target, and other objects in the scene, visualized by bounding boxes. B) Grasping task trial demonstration. C) Results of the experimental validation. Orange color indicates experts, blue naïve participants. D) Detection percentage during the navigation progress for successful and failed trials. As all trials had a different number of frames, they were normalized and afterwards binned into 50-step intervals.}
\label{fig_3}
\end{figure}

For the detection of objects, a pre-trained large YOLOv5 model with an image resolution of 1280 pixels showed the best retraining results for a batch size of 32 and 130 epochs, with a precision of 0.76, a recall of 0.65, a mAP of 0.70 and an inference time of roughly 1600 ms per frame during training due to the large model and image size. For the hand detection network, a small YOLOv5 model with an image resolution of 640 pixels for inference after training for 300 epochs with a batch size of 32 showed the best performance. Ultimately, the model had a precision of 0.98, a recall of 0.97, and an mAP of 0.98. For a comparison of all training runs, see Supplemental Files (Table S2). The object detectors perform at roughly 40 frames per second during the grasping task below. However, this value changes depending on the number of objects in the scene and the hardware used.

\subsubsection{Grasping Task}

In the grasping task, only the object and hand detectors were used without the object tracker and depth estimator. Experts grasped the target object successfully in 20 out of 20 trials (100\%), while naïve participants successfully grasped the target object in 10 out of 20 trials (50\%; Figure 3C). A failed trial occurred when the participant missed the target during their grasp, for example, due to overshooting. This results in an average performance of 30 out of 40 trials (75\%). Furthermore, the participants completed the trials successfully in a reasonable time despite getting no instruction to finish the task as fast as possible (M = 10.26s, SD = 6.79s). Interestingly, the naïve participants in this case were both faster and more consistent in their navigation speed (M = 9.74s, SD = 6.43s) than experts (M = 10.52s, SD = 7.12s).

While the average detection percentage (the number of frames in which the target object was correctly detected by the object detector divided by the total number of frames during the navigation part of the trial) in successful trials was 88.8\%, in failed trials, the objects were detected 83.4\% of the time on average (Figure 3D). For successful trials, the detection percentage stays consistent, while in failed trials a considerable drop in detection percentage can be identified after roughly two-thirds of a trial, reaching a minimum detection percentage of 39.4\%, while the smoothed average stays consistent at roughly 72\%. Validating this drop in detection percentage using the videos from the experiment reveals that the rotation of the camera and slow occlusion of the target by hand are two main factors for this drop in a total of three failed trials. At the end of a trial, the detection percentage reaches roughly 100\% before the grasping signal is sent. Importantly, causality between the object detection percentage and the success of a trial cannot be inferred, that is, it is not clear whether specific trials are not successful because the object detector had worse performance, or whether the detection percentage is lower in those trials because the participants moved their head too much. However, for successful trials, these numbers indicate reliable object detection that forms the base for the performance with the HANS.

\subsection{Tracking a target among similar objects}
\subsubsection{Object Tracking}

One component that adds to more robust guidance toward a target object is object tracking. More specifically, object detections from the scene are fed to the tracker, which performs operations over multiple frames to identify and track varying objects by assigning an identifier to each detection (Figure 4A). This is especially important for guiding the hand towards a target object, even when it is occluded for a number of frames, by keeping the object track active, using a specific identifier. It is then tracked to avoid jumping between multiple potential target objects of the same class.

\begin{figure}[b]
\centering
\setlength\fboxsep{0pt}
\setlength\fboxrule{0.25pt}
\fbox{\includegraphics[width=\linewidth]{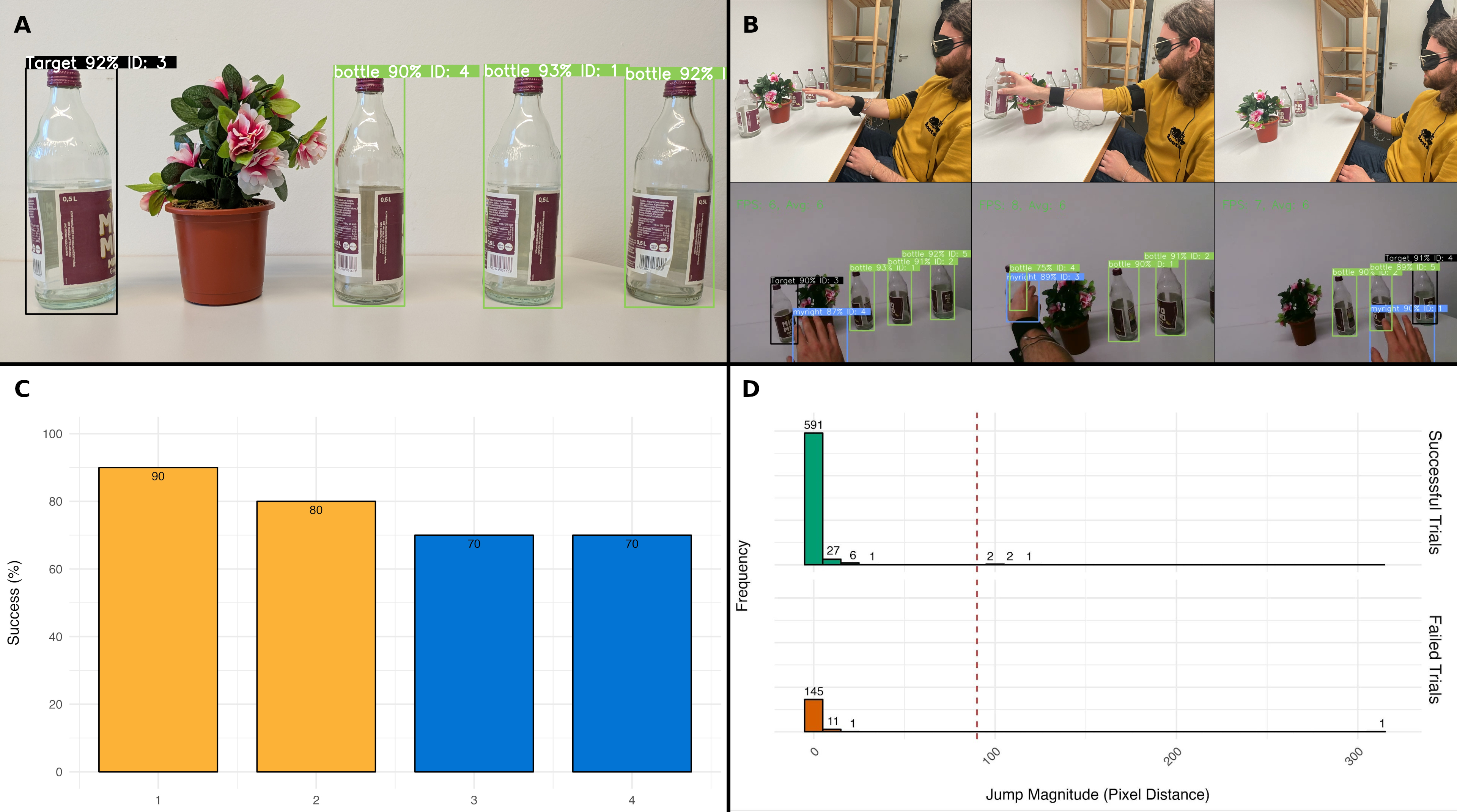}}
\caption{A) The target object is tracked amongst multiple instances of the same class using an identifier. B) Multiple objects task trial demonstration. C) Results of the experimental validation. Orange color indicates experts, blue naïve participants. D) Distribution of magnitudes of a tracking jump from one target object to a new one. The red line indicates the threshold for a jump occurring from one target to another (90 pixels). The histogram bins are 10 pixels wide.}
\label{fig_4}
\end{figure}

The HANS uses the StrongSORT algorithm \cite{Du2023} to track objects. This tracker consists of the OSNet re-identification network \cite{Zhou2019}, \cite{Zhou2021} that extracts image features to assign identifiers to detected objects. More specifically, the tracker is initialized with a nearest neighbor distance metric to track association between objects using the obtained object features from the re-identification network. Additionally, a Kalman filter is used to predict the object's position in the next frame, also considering the object’s constant velocity. Therefore, the tracker can filter object trajectories in the image space. We use pre-trained weights for the re-identification network from training on the Market-1501 dataset for person re-identification \cite{Zheng2015}, \cite{Lin2019}. We did not perform any evaluation since we did not compare tracking algorithms, as StrongSORT is the most recent advancement of DeepSORT and is the leading position in tracking algorithms. Therefore, we validated the tracker directly through usage in the project environment. The object tracker on top of the object detectors performs at roughly six frames per second during the multiple objects task. This value changes depending on the number of objects tracked and the hardware used.

\subsubsection{Multiple Objects Task}

In the multiple objects task, in addition to both the object detector and hand detector, an object tracker is utilized to identify different objects of the same class. Overall, the participants managed to perform the task successfully in 31 out of 40 trials (77.5\%). More specifically, experts performed the task successfully in 17 out of the total 20 trials (85\%), while naïve participants successfully grasped the target in 14 out of 20 trials (70\%; Figure 4C). Compared to the grasping task, participants performed faster in the multiple objects task (M = 6.10s, SD = 5.68s). Furthermore, the experts were faster and more consistent in their navigation speed (M = 4.58s, SD = 2.22s) compared to naïve participants (M = 7.96s, SD = 7.84s).

Thanks to the addition of the object tracker, the object detection percentage consistently reached 100\% during the navigation duration of the trial, that is, from navigation onset until the grasping signal was sent. Furthermore, as this task emphasizes the ability of the HANS to track only one specific target object among multiple objects of the same class, we report the magnitudes of target object jumps, i.e. shifts in the horizontal position (Figure 4D). In total, the majority of jumps were only minor, shifting the target position horizontally with a magnitude between zero and 10 pixels, a value corresponding to roughly 1.4 degrees of visual angle (minor fluctuations related to camera movements, indicating a constant focus on the target object), for both successful (93.8\%) and failed trials (91.8\%). Only five jumps in the successful trials, and only one jump in the failed trial, surpassed the critical threshold of 90 pixels (12.4 degrees of visual angle). Importantly, while this threshold is relatively arbitrary, choosing an even more conservative threshold does not alter the results as these jumps negatively impacted the outcome only for one trial with a jump magnitude of 313.5 pixels (43.1 degrees of visual angle).

\subsection{Avoiding Obstacles in the Environment}
\subsubsection{Depth Estimation}

The last component that is implemented to enhance the HANS is depth estimation which provides a depth map of the scene every few frames (Figure 5A) used to detect obstacles that lie within the direction of and are closer to the camera than the target object. It does so by guiding the hand in front of the obstacle in the direction of the target, or above the obstacle, depending on its orientation.

To choose a reliable depth estimator for the HANS, we evaluated and compared a set of depth estimator families that perform monocular single-shot depth estimation \cite{Patzold2025}. The depth estimators either performed relative depth estimation, predicting the relative disparity magnitudes between image pixels, or metric depth estimation, where predicted depth values are mapped onto an interpretable metric scale. To compare the different estimators, the speed, accuracy and robustness were measured on the HaND dataset \cite{HAND2024} to calculate a composite performance score, ranked the depth estimators. The best performing relative depth estimator was the MiDaS V2.1 model \cite{Ranftl2022} (see Figure S2A in Supplemental Files) with a median inference time of 36 ms per image and a median symmetric mean absolute percentage error of 0.073. Further, the best metric depth estimator was the small UniDepth model \cite{Piccinelli2024} (see Figure S2B in Supplemental Files) with a median inference time of 192 ms per image and a median absolute relative error of 0.174 m. Depending on the need for depth estimation in later stages of the project, a choice of relative or metric depth estimation can be made.

\subsubsection{Depth Navigation Task}

In the depth navigation task, both object detectors and an additional depth estimator were used, the latter of which periodically predicted a depth map that the HANS used to guide the participants’ hands around an obstacle towards a target object. Importantly, in this task, the object tracker was not used as only one target object was present, and the system’s performance should not be slowed down unnecessarily. Overall, 35 out of 40 trials were successful (87.5\%), with experts having a higher success percentage (95\%) than naïve participants (80\%; Figure 5D). The detection percentage of the target object was 89.4\% in successful trials and 100\% in failed trials. As the focus of this task was to avoid the obstacle in the line of grasping, we did not record or analyze any depth estimations. Furthermore, the navigation durations in the depth navigation task were larger than both former tasks combined, which can be explained by the additional cautious guidance in depth (M = 18.76s, SD = 14.08s). In this task, the experts were faster and more consistent in navigation duration during successful trials (M = 17.65s, SD = 8.60s) than naïve participants (M = 20.08s, SD = 18.91s).

\begin{figure}
\centering
\setlength\fboxsep{0pt}
\setlength\fboxrule{0.25pt}
\fbox{\includegraphics[width=\linewidth]{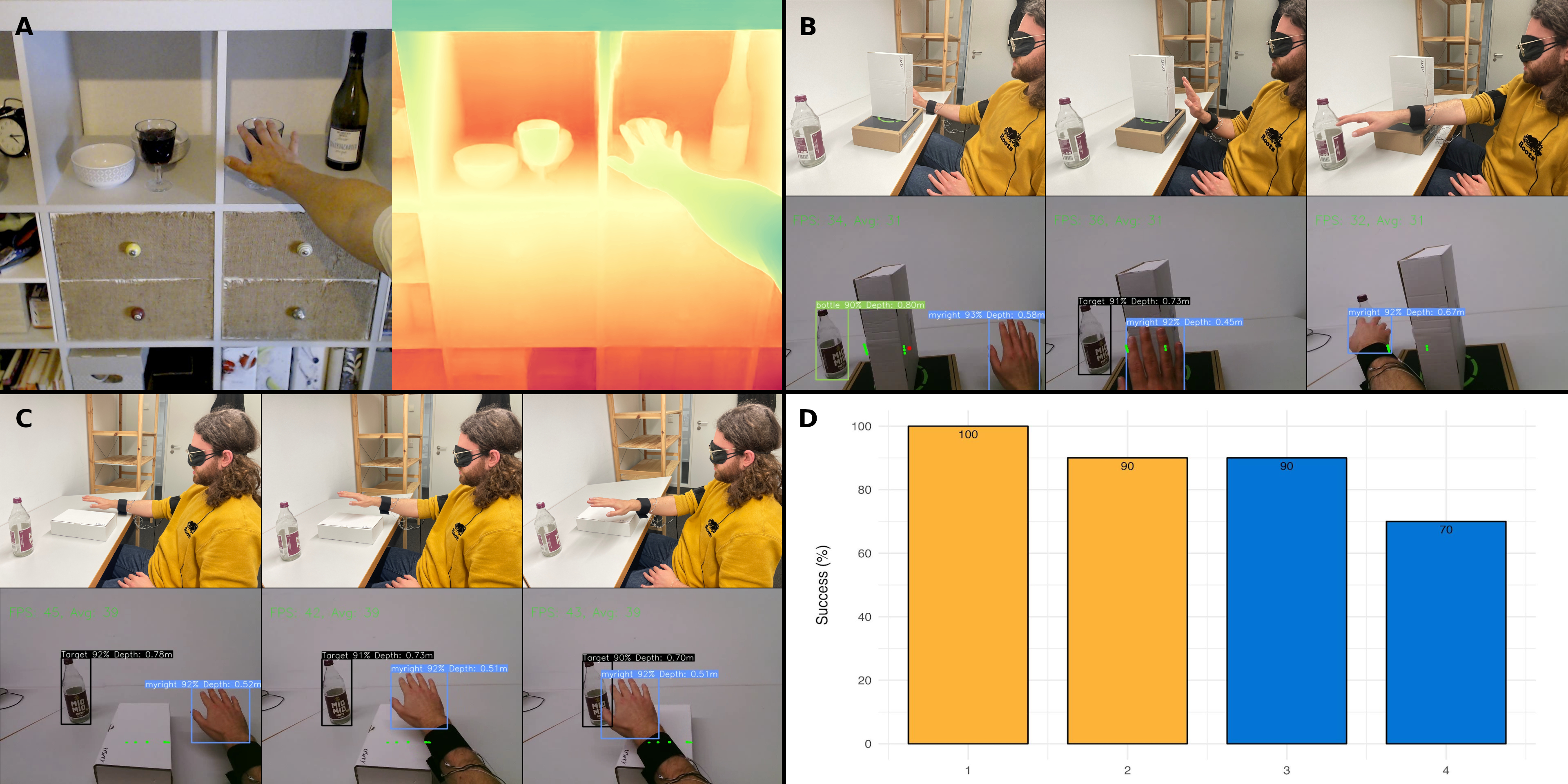}}
\caption{A) Depth map prediction for an example scene using the depth estimator. B) Depth navigation task trial demonstration with an obstacle that blocks horizontal movement. C) Depth navigation task trial demonstration with an obstacle that does not block horizontal movement. D) Results of the experimental validation. Orange color indicates experts, blue naïve participants.}
\label{fig_5}
\end{figure}

\subsection{Questionnaire}

After performing all three experimental tasks, the participants filled out a questionnaire with closed as well as open questions, the answers of which we use to assess the subjective evaluation of the tactile bracelet, the tasks and the HANS. More specifically, the closed questions formed categories that enable this evaluation (see Table S1 in Supplemental Files). The participants felt comfortable and had a positive impression of using the bracelet (discomfort: median = 1.0, mean = 1.25, SD = 0.46; engagement: median = 4.0, mean = 4.38, SD = 0.52). They reported no problems interpreting the vibrations (interpretability: median = 4.0, mean = 3.63, SD = 0.74; vibrations: median = 4.5, mean = 4.50, SD = 0.54) and rated the training positively (training: median = 5.0, mean = 4.38, SD = 1.06). In line with ratings (Figure 6), one participant noted that “using calibrated vibration intensities made wearing the bracelet more comfortable”. Another participant mentioned that “vibrations were quite easy to distinguish”, but noted that “if I were to use it longer perhaps at some point my skin would get numbed”, showing a future need to focus on longer-term usage effects.

\begin{figure}[t]
\centering
\includegraphics[width=\linewidth]{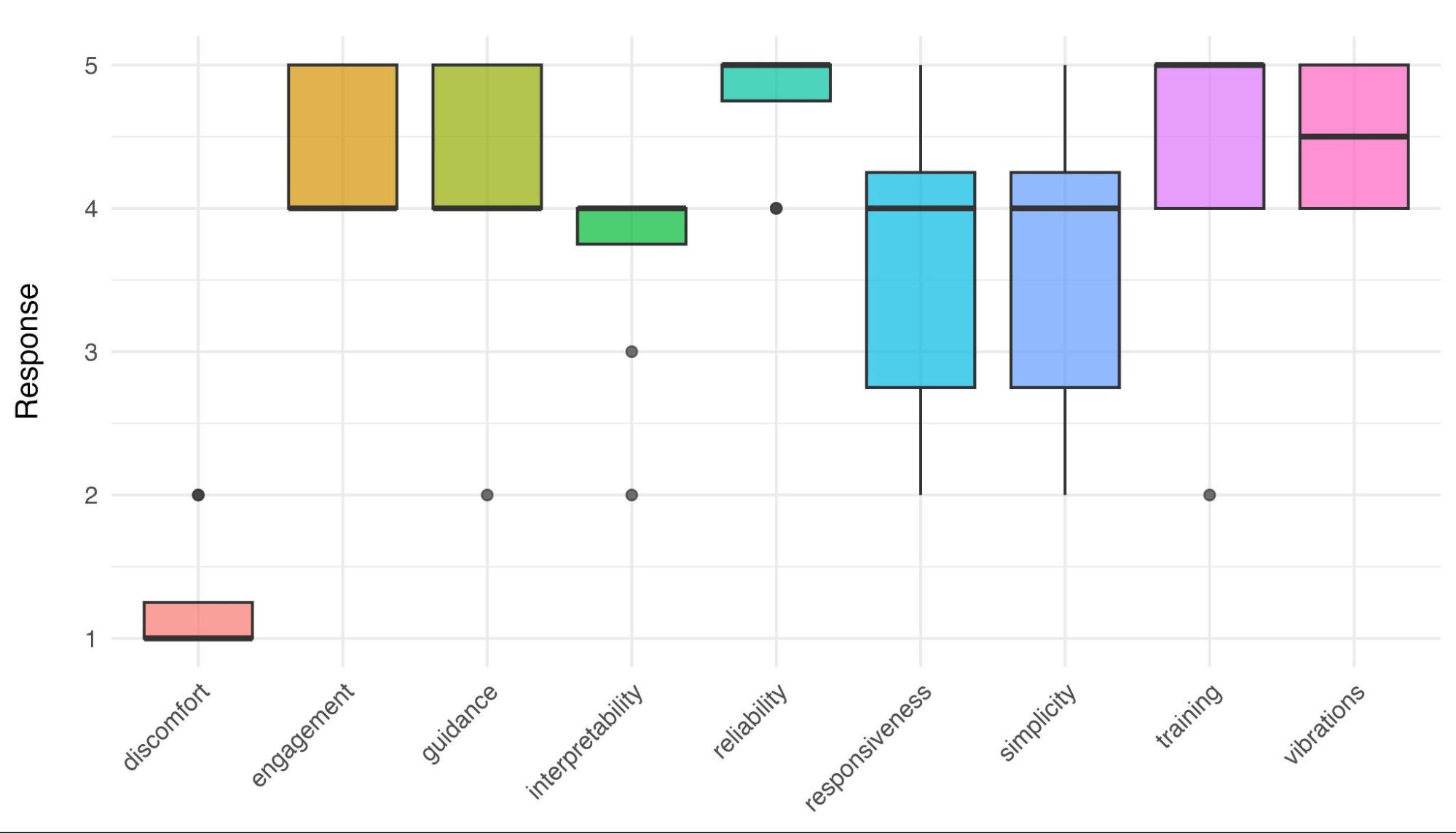}
\caption{Distribution of the responses to the questionnaire for all topics aggregated across participants.}
\label{fig_6}
\end{figure}

During the experiment, participants felt vibration commands were guiding them in a predictable, consistent manner (guidance: median = 4.0, mean = 4.13, SD = 0.99), and that the bracelet was generally responsive (responsiveness: median = 4.0, mean = 3.63, SD = 1.19). However, one participant pointed out that it can be “confusing to differentiate the top with right and left“ due to a small wrist size. Nonetheless, all participants appreciated that they could rely on the bracelet to perform the task (reliability: median = 5.0, mean = 4.75, SD = 0.46). This key aspect is confirmed by their responses in the open section, where they mentioned that “it's pretty good experience to know that I can grasp targeted object with closed eyes and rely on the tactile bracelet” and “[it] is pretty good that I can avoid the obstacle and don't hit it to grasp the targeted object”. Importantly, one participant noted that “trying to rely on the bracelet and switching off intuition helps navigation”, indicating that the system can be used intuitively and reliably. Overall, participants rate the tasks as doable thanks to the bracelet (simplicity: median = 4.0, mean = 3.63, SD = 1.19).

Additional remarks from the open section were related to the limitations of the camera (“sometimes the camera cannot see the targeted object and I have to move my head”), especially in the depth navigation task (“I felt like I had to move my head more to make everything fit in the FOV of the camera”, “last task was most fun but was a little annoying when hand got lost because of camera field of view”). Moreover, several minor suggestions related to how training and calibration procedures could be improved were proposed.

\subsection{Evaluation with Blind Participant}

To assess the system's usability for the target user’ group in more natural contexts, we planned an evaluation session with the blind community. First, we invited two late blind participants for an initial bracelet testing session in a café. The aim was to receive critical feedback for further development of the HANS. Importantly, a new type of navigation that differed from the one used with the blindfolded participants was suggested. More specifically, the hand would be navigated to a target object by first guiding alongside the horizontal axis only, followed by guidance across the vertical axis, enabling a simpler interpretation of the direction commands.

\begin{figure}[b]
\centering
\includegraphics[width=\linewidth]{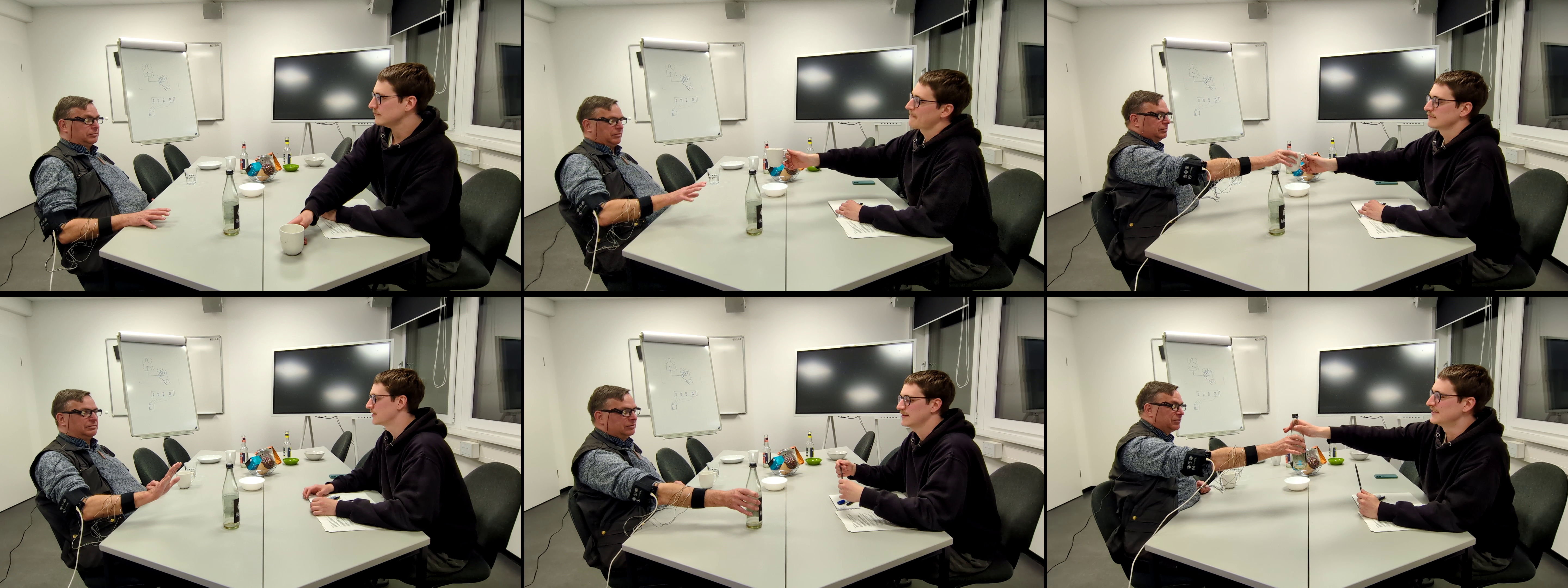}
\caption{Interaction between blind participant and experimenter. In the upper row, the experimenter picks the object the participant grasps above the table. In the lower row, the participant picks the object and passes it to the experimenter.}
\label{fig_7}
\end{figure}

As a follow-up, we invited one of the participants for the evaluation session. The session was conducted within a less structured context than the experiment with blindfolded participants, aiming to validate the system's robustness in less sterile scenarios while still being able to control for some conditions, like the lighting. One of the experimenters sat across the participant with their hands on the table, providing a distractor for the hands detector throughout the whole session. After self-paced training, the participant performed an adjusted variant of the grasping task, that is, 10 trials with three different objects placed in multiple starting positions within a reaching distance, on an office table in front of the participant. Next, we continued with an interaction task, during which the experimenter and participant grasped one of the objects from their side of the table and passed them to each other (Figure 7). From the participant’s point of view, they either had to perform a regular grasping trial to pass the object to the experimenter consequently, or had to grasp an object held in the air across the table by the experimenter. Both the participant and experimenter were supposed to pass three objects to each other, resulting in a total of six interaction trials. Finally, the participant gave general remarks about the experiment, the tactile bracelet, and the HANS in a questionnaire.

The participant successfully completed both tasks, with an accuracy of 9 out of 10 grasped objects in a grasping task (90\%) and 5 out of 6 in the interaction task (83.3\%). Importantly, the cause of the failed trials was grasping next to the target upon receiving the grasping signal in both tasks. Furthermore, the navigation duration describes the time point when the hand and target are both visible in the frame until the grasping signal is sent. The average navigation duration in the grasping task was 35.8s and 7.6s in the interaction task.

Similarly to the blindfolded participants, the overall feedback from the blind participant was positive. The participant mentioned that the training session enabled a fast understanding of how to interpret the vibration signals to use the tactile bracelet in the experimental tasks successfully. Furthermore, the participant also commented on potential shortcomings and the possibility of designing vibration patterns differently. For instance, when in a trial the target would be occluded by the hand without correctly sending the grasping signal, but sending no signals instead, the participant commented that this could also be seen as a grasping signal. Therefore, in the following trials, the participant integrated their available knowledge also to interpret this “alternative” grasping signal, showing how even complex scenarios can be accounted for by learning how to interpret the vibration signals to ultimately use the tactile bracelet successfully. Furthermore, the participant proposed adding the pulsating vibration pattern of the top motor instead of a continuous one, and of the alternating vibration pattern for diagonal directions. These suggestions highlight significant individual user differences in how vibration signals are perceived and, consequently, how vibration patterns could be designed depending on the needs of the tactile bracelet’s user.

\section{Discussion}

In this study, we validated the HANS for the tactile bracelet to allow autonomous grasping movements. The automated system for navigating the hand comprises three main modules, including a pair of object detection models, an object tracking algorithm, and a depth estimator. Together, these modules allow for the reliable conversion of visual cues in the environment to the guiding signal, which is subsequently sent to the tactile bracelet and leads the user’s hand towards a target object. The impact of the HANS has been tested by developing a paradigm consisting of three complex scenarios of grasping objects in varying surroundings, including distractors or obstacles, that resemble everyday life. These scenarios included the search for a target object between multiple distractor objects of different or the same category, or when confronted with obstacles in the grasping path. In particular, both blindfolded participants in the laboratory, as well as blind participants in a café used the tactile bracelet with the HANS to grasp target objects in such scenarios. Our study's results highlight the tactile bracelet's usability for the blind community. Thus, the tactile bracelet with the HANS qualifies as a meaningful tool for facilitating an autonomous life for blind and visually impaired individuals.

Given these achievements, it is worthwhile to mention the most prominent shortcomings of the tactile bracelet system. Firstly, the current speed of the system is a bottleneck restricting its functionality for real-world applications. While in the experimental setup, the object tracker and depth estimator were either turned off or on for the whole duration of the task, and the end-user of the system would need to be able to use all of the components to guarantee its optimal performance. A potential solution for that problem would be the dynamic contextualization of the system activity based on the scene composition. For example, if, based on an initial assessment of the depth estimator, no obstacles are present in the scene, it could be either turned off or run less frequently. Alternatively, if several obstacles are detected, the depth estimator could be run on more frames. However, while this idea optimizes the use of available computational power, the selection of the hardware is essential. One option would be to utilize smartphones' vast popularity and growing computing capacities, a feature that enables their effective usage as assistive technology \cite{Khan2021}, \cite{Pundlik2023}, \cite{Senjam2021}. Another option is to move the computational load of the navigation system to the cloud and dynamically interact with it based on the web requests, transforming the tactile bracelet system into an on-demand assistive device within the Internet-of-Things \cite{Baucas2021}, \cite{DeFreitas2022}, \cite{Rajarajan2024}. Regardless of the choice, each option would vastly enhance the system's scalability.

Secondly, hardware limitations restrict the capacity of the system. The current version of the bracelet limits the number of signals that can be sent and, consequently, processed by users. Therefore, although the current navigation algorithm incorporates information about the depth properties of the scene, it is limited to sending guiding signals only in two dimensions at once using the four vibration motors, restricting the possibility for the user's grasping trajectory to resemble that of normally sighted people. Presently, we are working on creating new variants of the bracelet, with one design – a sleeve with horizontal and vertical vibration motors spread along the forearm – already tested within the scope of a project-related Master’s thesis \cite{Christiadi2025}. Nonetheless, further development in that area should follow, with associated modifications to the navigation logic to fully incorporate any new available capabilities.

Lastly, the current version of the system is limited by the use of text input to determine the target object. Thus, even though the experimenter is removed from the navigation loop, the system's usability is still limited as it can be utilized only in a particular context. The development and addition of the interactive natural language processing module that will handle users’ auditory requests to determine both the context of the situation and the target objects is a crucial next step to make the system fully autonomous. For example, if the user visits a supermarket to buy specific fruits, the system should navigate to any target object that is in the shopping list unless a specific order is defined. However, if the user prepares a meal and provides information about subsequent elements they want to add to their dish, the system should search for and navigate toward target objects sequentially. Additionally, the system should be able to provide feedback to interact with the user continually. Such a system could be based on large language models \cite{Naveed2023} that can process natural language prompts \cite{Kaddour2023}, \cite{Xie2023}. We are currently working on introducing that component as it is an essential next step of the project development. Nonetheless, notwithstanding the above limitations, this HANS is the initial reliable solution and has great potential to pave the way to more independence in the daily lives of blind people.

Our solution contributes significantly to the growing field of task-specific assistive devices. The tactile bracelet system enables its user to grasp independently without blocking the auditory channel, the most important source of information for blind people \cite{Roder1999}. While several previous devices utilized tactile signals to scan the environment and navigate the user’s hand \cite{DePaz2023}, \cite{Satpute2020}, \cite{Yu2016}, our developments provide novel functionality enabling navigation towards specified target objects. Significantly, our HANS could be adapted for use with other already existing and newly developed devices aiming to help with grasping. Thus, our findings provide exciting opportunities for the future, with the promise of scalability within the scope of our project and in their potential integration with other solutions.

The tactile bracelet system poses interesting questions within the scope of the general field of assistive devices. As we based our research on the principle of sensory augmentation \cite{Macpherson2018}, we assume that continuous usage of the bracelet might create new sensorimotor contingencies \cite{ORegan2001}. However, as our device is task-specific and not of general purpose, it would likely be used only whenever needed rather than at all times. Thus, while it is possible that prolonged usage would enhance the ability to detect and process tactile stimuli or change perception of the surrounding space, the current knowledge does not enable the formulation of a specific hypothesis. Therefore, further longitudinal studies aimed at analyzing the long-term usage effects of our device would be of great interest.

As the final goal of our device is to help the blind population, its validation with the users from the target group was essential. Results of our visit to a cafe with two visually impaired participants are positive and show that the bracelet could be used independently in the future. However, it is worth noting that the blind population is not homogeneous, with major differences between congenitally and late blind people \cite{Bedny2012}, \cite{Collignon2013}. In relation to several aspects of the tactile bracelet system, it has been shown that a type of blindness affects how those affected process tactile stimuli \cite{Wan2010}, auditory stimuli \cite{Voss2016}, and their surroundings \cite{Cattaneo2008}, \cite{Crollen2012}, \cite{Gori2020}, \cite{Schinazi2016}, impacting how they interact with the world. Therefore, the usability of the tactile bracelet system should be assessed for each user. Nonetheless, our findings from the validation within the target group highlight the potential benefit for the blind community. 

In summary, our study shows that the addition of the HANS enables efficient processing of visual inputs to tactile signals and, as an effect, autonomous usage of the tactile bracelet as an aid to grasp target objects. Thus, it contributes significantly to the field of assistive technology, increasing the potential for easier access and more autonomous usage of a much-needed aid for blind people \cite{WHO2024}. Overall, our system serves as a landmark development that could help assist blind people in grasping and potentially be adapted for other specific tasks, providing the visually impaired with higher levels of independence and enhancing their quality of life.

\section{Conflicts of Interest}

Author Silke M. Kärcher is the CEO of feelSpace GmbH. The remaining authors declare that the research was conducted without any commercial or financial relationships that could be construed as a potential conflict of interest.

\bibliography{bibliography}
\bibliographystyle{IEEEtran}

\end{document}